\documentclass[fleqn,usenatbib]{mnras}

\usepackage{newtxtext,newtxmath}

\usepackage[T1]{fontenc}
\usepackage{ae,aecompl}

\usepackage{mathtools}
\usepackage{amssymb}
\usepackage{graphicx}
\usepackage{natbib}
\usepackage{color}

\usepackage{graphicx}	
\usepackage{amsmath}	
\usepackage{amssymb}	

\newcommand{\arcsecond}[2]{\ensuremath{#1''\!\!.#2}}
\newcommand{\arcminute}[2]{$#1'\!\!.#2$}
\renewcommand{\micron}{\ensuremath{\mu\text{m}}}
\newcommand{\angstrom}{\ensuremath{\text{\AA}}}

\newcommand{\sbB}[1]{$\mu_0({\rm B}) #1$}
\newcommand{\sfr}[1]{$#1\rm\,M_\odot\,yr^{-1}$}

\newcommand{\mstar}[1]{\ensuremath{{\rm M}_\bigstar#1}}
\newcommand{\lmstar}[1]{\ensuremath{{\rm log\left(\,M_\bigstar\right)}#1}}
\newcommand{\zstar}[1]{\ensuremath{{\rm Z}_\bigstar#1}}
\newcommand{\zsol}[1]{\ensuremath{{\rm #1\,Z}_\odot}}

\newcommand{\about}[1]{\ensuremath{\sim\!\!#1}}

\makeatletter
\newcommand{\Rom}[1]{\expandafter\@slowromancap\romannumeral #1@}
\makeatother

\newcommand{\ionl}[3]{[#1$\,${\small \Rom{#2}}]\relax$\lambda$#3}

\newcommand\HB{{\rm H}$\beta$}

\newcommand{\HI}{H{$\,$\footnotesize I}}
\newcommand{\HII}{H$\,${\small II}}

\newcommand\nobrkhyph{\mbox{-}}
\newcommand{\pegase}{\textsc{P\'egase}}
\newcommand{\gband}{g\nobrkhyph{}band}
\newcommand{\vp}{VIRUS\nobrkhyph{}P}

\title[Recent Starburst in an LSB]{A Recent Starbust in the Low Surface Brightness Galaxy UGC~628}

\author[J. E. Young, R. Kuzio de Naray, and Sharon X. Wang]{
  J.~E.~Young,$^1$
  Rachel Kuzio de Naray,$^2$
  Sharon X. Wang$^3$\\
  $^1$Department of Astronomy, Mount Holyoke College, 50 College Street, South Hadley MA, 01075 jyoung@mtholyoke.edu.edu\\
  $^2$Department of Physics \& Astronomy, Georgia State University, P.O.\ Box 5060, Atlanta, GA 30302-5060 kuzio@astro.gsu.edu\\
  $^3$Department of Terrestrial Magnetism, Carnegie Institution for Science, 5241 Broad Branch Road, NW, Washington, DC 20015, USA
}

\date{Accepted XXX. Received YYY; in original form ZZZ}

\pubyear{2019}

\begin{document}
\label{firstpage}
\pagerange{\pageref{firstpage}--\pageref{lastpage}}
\maketitle

\begin{abstract}
We present the star-formation history of the low surface brightness (LSB) galaxy UGC~628 as part of the MUSCEL program (MUltiwavelength observations of the Structure, Chemistry, and Evolution of LSB galaxies). The star-formation histories of LSB galaxies represent a significant gap in our knowledge of galaxy assembly, with implications for dark matter / baryon feedback, IGM gas accretion, and the physics of star formation in low metallicity environments. Our program uses ground-based IFU spectra in tandem with space-based UV and IR imaging to determine the star-formation histories of LSB galaxies in a spatially resolved fashion. In this work we present the fitted history of our first target to demonstrate our techniques and methodology.  Our technique splits the history of this galaxy into 15 semi-logarithmically spaced timesteps. Within each timestep the star-formation rate of each spaxel is assumed constant. We then determine the set of 15 star-formation rates that best recreate the spectra and photometry measured in each spaxel. Our main findings with respect to UGC~628 are: {\bf a)}  the visible properties of UGC~628 have varied over time, appearing as a high surface brightness spiral earlier than 8~Gyr ago and a starburst galaxy during a recent episode of star formation several tens of Myr ago, {\bf b)} the central bar/core region was established early, around 8-10 Gyr ago, but has been largely inactive since, and {\bf c)} star formation in the past 3~Gyr is best characterised as patchy and sporadic.
\end{abstract}



\section{Introduction}
\label{sec:intro}
\newcommand{\figref}[1]{Figure~\ref{fig:#1}}\
\newcommand{\secref}[1]{Section~\ref{sec:#1}}\
\newcommand{\tblref}[1]{Table~\ref{tbl:#1}}\

The modern picture of galaxy assembly is effective at explaining a wide range of galaxy classes. Star formation is fueled by cool gas, and spiral discs generally progress from late to early types as the gas is depleted. Ram-stripping can accelerate this process, while interactions and mergers are responsible for the formation of bulges and ellipticals. However, this picture is heavily slanted toward explaining galaxies which are easy to observe. With central surface brightnesses \sbB{>22}, fainter than typical sky backgrounds, low surface brightness (LSB) galaxies are less studied than their high surface brightness (HSB) counterparts and often chronically underrepresented in surveys.

The term ``Low Surface Brightness Galaxy'' has recently evolved into an umbrella term, encompassing a wide range of physically distinct galaxy classes. For example, some of the most extreme LSB galaxies are the ultra-diffuse cluster galaxies \citep[e.g.,][]{vanDokkum2015}, which are red/dead discs likely ram-stripped of cool gas billions of years ago. The most numerous LSB galaxies are undoubtedly gas-poor dwarf galaxies, which are thought to have expelled their gas in an early burst of star-formation due to their low escape velocities \citep[e.g.,][]{Sawala2010,Hopkins2012}. Dark matter-poor structures such as the Leo Ring \citep{Schneider1983} and the somewhat contested ``ghost galaxy'' \citep{vanDokkum2018}, which are gas rich but late-forming due to long dynamical times, have also joined the ranks of LSB galaxies. For each of these classes there exists a clear and unique explanation as to why their stellar populations are so sparse.

In contrast, our work focuses specifically on gas rich, blue, dark matter-dominated, rotationally supported LSB disc galaxies, henceforth referred to as ``LSB spirals''. To be clear, though, the division between LSB and HSB spirals is somewhat artificial, with no discernible bimodality in surface brightness \cite[e.g.,][]{McGaugh1995distribution}; our targets are simply the faint tail of the same distribution which includes the Milky Way. For this class of objects there is no obvious reason why more of their gas has not yet been turned to stars. Indeed, there may be multiple mechanisms to arrest their star formation.

The abundant gas in the discs of LSB galaxies typically exhibits low densities and low metallicities. A straightforward application of the Schmidt-Kennicutt Law predicts little/no star formation in LSB discs, but this only pushes the question back a step; why does the gas remain at low densities? Aside from exceptional cases, such as NGC~4395 \citep{Heald2008}, it is unlikely that the \HI{} gas in most LSB spirals is recently accreted material. More likely, the majority of the gas in the majority of LSB spirals was accreted at early times, and the gas has remained in a stable but diffuse disc since.


One approach is to use the assembly information fossilized into the stellar populations to paint a clearer picture of the formation of LSB spirals. With average colours somewhat bluer than HSB spirals \citep{McGaugh1994,deBlok1995}, LSB spirals would seem to be young, unevolved systems, however early studies quickly showed that this simple explanation was insufficient. For example, \cite{Zackrisson2005} use multi-band photometry to show that LSB galaxies are poorly modeled by only a young stellar component, and have likely been forming stars for quite some time.

A more viable explanation is that most LSB spirals are indeed old, but their star formation has been patchy and sporadic. \cite{Boissier2008} suggest that the red FUV-NUV but blue optical colours exhibited by many LSB spirals are best explained by a low and slowly evolving star formation rate punctuated by intense bursts of star formation. The few red LSB spirals in their sample, then, have simply had a longer quiescent period since their most recent burst. Some issues remain; for example, the bursts required to replicate the observed properties of LSB spirals are rather extreme, up to hundreds of \sfr{} for the most massive LSB spirals and typically tens of \sfr{} for LSBs spirals in the Milky Way mass range. Also, \cite{Boissier2008} point out that it is difficult to rule out the possibility that the red FUV-NUV colours are the result of a truncated IMF. In this alternate scenario where the IMF is truncated, the star formation rate in LSB spirals is high, but there are insufficient high mass stars to produce the expected hard UV photons, resulting in red UV colours and inaccurately low emission line-derived star-formation rates. However, caveats aside, the ``patchy/sporadic'' mode of star formation described in \cite{Boissier2008} remains highly plausible, and effectively explains many of the observed properties of LSB spirals.

Likewise, \cite{Vorobyov2009} use hydrodynamic models to show that LSB spirals are likely many Gyr old and dominated by patchy/sporadic clumps of star formation. \cite{Schombert2014} model the optical and IR colours of LSB galaxies and rule out the possibility that typical LSB galaxies are more the 5~Gyr younger than typical HSB galaxies. They find that the driving difference is a low overall star-formation rate with sporadic bursts to help explain the variation in colours, versus the initially fast but now declining star-formation rates of HSB galaxies.

Metallicity may play a role in regulating the mode of star formation: Using an N-body simulation, \cite{Gerritsen1999} find that star formation is suppressed in LSB spirals primarily as a result of low metallicities. With lower cooling efficiencies, the formation of molecular clouds is slower and rarer. These simulations make predictions in-line with patchy/sporadic star formation. In their simulations the star-formation rate has strong fluctuations, however the amplitude of the fluctuations are lower and the durations of the quiescent periods are longer than those derived from FUV-NUV colours by \cite{Boissier2008}. Likewise, \cite{Gnedin2011} note that, in most environments, the shielding of H$_2$ is partly accomplished by dust, which is rare in the low metallicity environments of LSB spirals.

The concept of sporadic, patchy, disc-wide star-formation is somewhat at odds with the inside-out description of galaxy evolution, wherein spiral galaxies form stars near their centres earlier and faster than in their outer discs. Flatter optical colour gradients \citep[e.g.,][]{Matthews1997,Galaz2006} and flat or inverted metallicity gradients \citep{deBlok1998I,Young2015} also seem to imply that evolution in LSB galaxies does not seem to show an inside-out or outside-in preference. However, \cite{Bresolin2015} find that this may be an effect of the larger scale radii of LSB Spirals; we will revisit this issue in \secref{context}. 

There are also limitations to the analyses to-date. For example, \cite{Zackrisson2005} and \cite{Kim2007} are unable to place hard constraints on the ages of LSB galaxies due to the degeneracy between age and star-formation history. Both of these works emphasise the patchy nature of star formation in LSB galaxies, but both rely on whole-galaxy measurements. Even with an accurate whole-galaxy star-formation history, open questions would remain, such as: have LSBs always had patchy star formation, or did they begin in a fashion similar to HSBs and then diverge? If star formation in LSB galaxies is best characterised by sporadic bursts, then what is the duty cycle of these bursts, and what are the visible properties of LSB galaxies over this cycle? Because LSB galaxies seem to have had star-formation histories that are variable both in time and across their discs, both spatial and spectral resolution are needed to resolve this issue.

 
In addition to providing insight into galaxy assembly, the early star-formation history of LSB galaxies may be a key element in resolving the ``cusp-core'' problem. A growing body of evidence suggests that the dark matter profiles of typical Milky-Way mass LSB galaxies are better fit by isothermal (cored) profiles rather than the cuspy NFW profiles produced by most cold dark matter simulations \citep{Borriello2001,deBlokBosma2002,KuziodeNaray2008}.

Supernova feedback is typically invoked as an explanation. In this scenario, a central star-forming episode expels enough gas from the centre of a galaxy to gravitationally drag a significant amount of dark matter from the centre, flattening a cusp into a core \citep[e.g.,][]{Pontzen2012,Governato2012}. This mechanism can explain the cored dark matter halos of dwarf galaxies, but a very significant amount of energy is required to restructure the halos of massive LSB galaxies; for example, \cite{DiCintio2014} find that the effectiveness of supernova feed back drops off for galaxies more massive than \lmstar{>8.5}. Resolving the star-formation histories of LSB galaxies allows for a direct comparison between the intensity of early central star formation and the energetics needed to reshape dark matter halo. Additionally, a clearer picture of LSB galaxies before and during the halo resculpting processes would provide a key test for this theory by allowing the identification of galaxies at different points in this process.


In order to address these questions, we present the MUSCEL program (MUltiwavelength observations of the Structure, Chemistry, and Evolution of LSB galaxies). In this paper we use optical IFU spectra in tandem with Spitzer IRAC and Swift UVOT observations to derive the spatially resolved star-formation history of the LSB galaxy UGC~628.  The techniques presented here will be used to similarly determine the histories of other galaxies in our sample. The structure of this paper is as follows: In \secref{obs} we outline our target selection and observations, including a description of the established physical characteristics of UGC~628. In \secref{analysis} we present our analysis, including star-formation history fitting. The results of our analysis, including a best-fit history and a discussion of its implications, are presented in \secref{results}. Finally, in \secref{summary} we summarise these results and discuss future directions for the MUSCEL program.

\section{Target Selection and Observations}
\label{sec:obs}

\subsection{Target Selection and Program Architecture}
The crux of our project is the comparison between the observed spectral energy distributions (SEDs) of LSB spirals and SEDs generated from synthetic star formation histories (SFHs). Our modeling program, described below in \secref{analysis}, uses the spectral synthesis program \pegase{} \citep{PEGASE}. In order to provide the maximum constraint on the SFH, our program uses ground-based optical spectra in tandem with space-based Spitzer 3.6\micron{} and Swift UVM2 photometry.

The UVM2 filter is a broad-band filter similar to Galex NUV. Although our modeling program does not derive properties from the UVM2 measurements directly, the UVM2 photometry constrains the current/recent star formation and helps constrain the extinction parameter. Our program was awarded approximately nine hours\footnotemark[1] to collect UV data on LSB spirals of interest to the MUSCEL project, as part of the Swift Cycle 10 Guest Investigator Program.

\footnotetext[1]{Proposal ID:1013267}

The 3.6\micron{} photometry complements the UVM2 data by constraining the integrated star-formation history. As with the relationship between the UMV2 data and the current star-formation rate, this constraint comes from the fact that the 3.6\micron{} photometry is correlated with stellar mass. We do not derive stellar mass directly prior to SFH fitting, nor do we assume any mass-to-light ratio. Our observations are drawn from archival Spitzer data (warm Spitzer data in the case of UGC~628).

Finally, we take advantage of the intermediate age indicators in the optical spectrum, such as the H$\delta$ line, the 4000\angstrom{} break, and the overall shape of the continuum, by including in our SEDs spatially-resolved optical spectra in the 3600-5700\angstrom{} range. Our spectral data were collected using the \vp{} IFU at the 2.7m Harlen J. Smith telescope. \vp{} is a fiber-fed IFU spectrograph with \arcsecond{4}{16} fibers which we use to construct spectral data cubes with a \arcsecond{0}{5} plate scale and a seeing resolution limit of 2\arcsec{} (see \secref{photovalid}). Again, we do not derive properties directly from the optical spectra, but the presence of these features allows our fitting method to constrain the star formation history at intermediate ages; see \cite{Young2015} (hereafter Paper~I) for a thorough discussion of the reduction methods and physical properties of UGC~628 derived directly from the optical spectra.

\begin{figure}
  \includegraphics[width=0.5\textwidth]{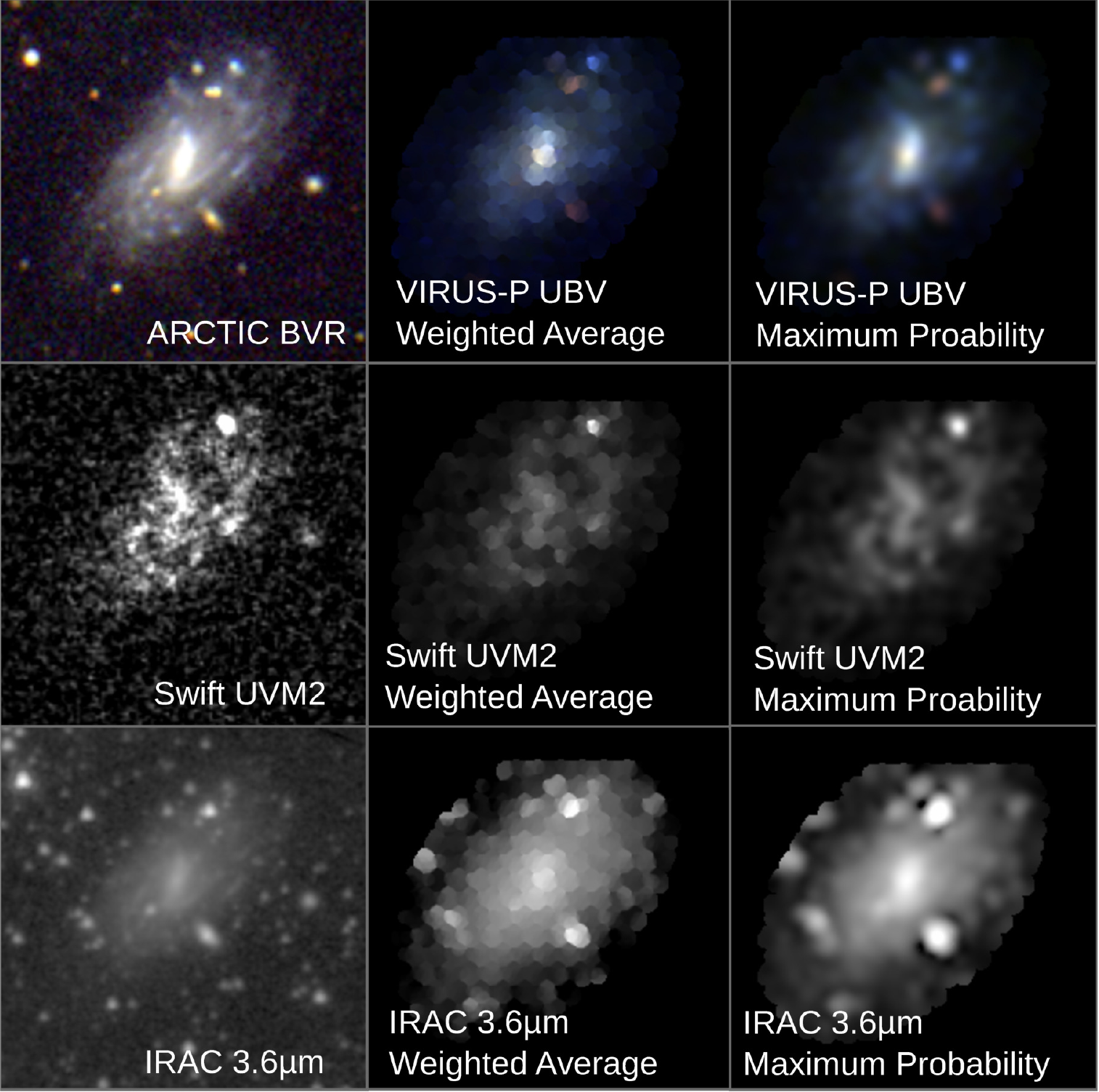}
  \caption{{\bf left:} UGC~628 seen with ARCTIC BVR, Swift UVM2, and Spitzer IRAC 3.6\micron{} filters. {\bf middle: } Weighted average reconstructed images from \vp{} data convolved with UBV filters ({\bf top}), or from matching synthetic fibers used to sample the corresponding image in the left column. {\bf right:} Maximum probability reconstructed images.}
  \label{fig:reconstruct}
\end{figure}

Below in \secref{photovalid} we discuss the reduction and analysis of the \vp{} data, involving the reconstructing the \vp{} image plane. As a comparison to our reconstructed \vp{} data, we include in \figref{reconstruct} BVR data taken with the ARCTIC imager on the Apache Point Observatory 3.5m telescope.

Our targets are drawn from the LSB galaxies cataloged in \cite{Kim2007} and \cite{McGaugh1994}. Our initial study includes only galaxies that fit comfortably within the \arcminute{1}{7}$\times$\arcminute{1}{7} \vp{} field-of-view while still spanning many 2\arcsec{} seeing-limited resolution elements (typical of Mt.~Locke). In practice, that means that the distances to our target galaxies are typically 10-100Mpc. To maximise galaxy coverage, minimise dust obscuration, and minimise the number of distinct stellar populations along each line-of-sight, we avoided galaxies that are edge-on ($i>85^\circ$). Finally, our target list was also restricted to objects with archival Spitzer IRAC 3.6\micron{} data. The first target in our program with completed ground-based spectra is UGC~628, which is the focus of this paper.

With a central B-band surface brightness of $23.1~\rm mag~arcsec^{-2}$ \citep{Kim2007}, UGC~628 falls clearly on the LSB side of the surface brightness continuum, though it is by no means an extreme member of the population. Using broad-band photometry \cite{Kim2007} estimate $\lmstar{=}$ 10.65 or 10.80, depending on the IMF assumed, making it a near match to the $\lmstar{} = 10.7$ for the Milky Way \citep{Flynn2006,McMillan2011}.

\subsection{Data Reduction and Image Reconstruction}
\label{sec:reduce}

Our program uses data from three instruments: The \vp{} IFU imaging spectrograph on the 2.7m Harlen J. Smith telescope, the UVOT imager on the Swift Space Telescope, and the IRAC imager on the Spitzer Space Telescope.

We developed a \vp{} reduction pipeline optimised for low surface brightnesses, detailed in Paper~I. Once reduced, the \vp{} spectra were mapped onto a spectral data cube with a \arcsecond{0}{5} plate scale and a 20\angstrom{} spectral resolution. The angular resolution in our spectral data cubes is limited partly by the seeing at Mt.~Locke (typically around 2\arcsec{}), but also by the \arcsecond{4}{16} diameter of the \vp{} fibers. Our observing strategy uses a 6-point dither pattern which allows us reconstruct the image plane from the fiber data at with an angular resolution of around 2\arcsec{}. We have chosen to reconstruct the image plane at a \arcsecond{0}{5} plate scale since adopting a plate scale equal to our 2\arcsec{} plate scale would significantly blur out resolution elements not aligned with the 2\arcsec{} pixel boundaries. As in Paper~I, we over-sample the angular resolution and recognise that any apparent structures smaller than 2\arcsec{} are not real.

The only modifications from the \vp{} data reduction procedures in Paper~I are as follows: {\bf 1)} Because our spectral synthesis program, \pegase{}, has a resolution of 20\angstrom{} in the optical part of the spectrum, the \vp{} data are binned at an early stage to 20\angstrom{}~bins aligned to the \pegase{} wavelength bins in the rest frame of UGC~628. By binning only once at this step, we avoid further degradation of the spectra which would be incurred in subsequent binning steps. {\bf 2)} Instead of a simple weighted average for image reconstruction, we now use a maximum probability image reconstruction, which minimises the residuals between fiber data and a Gaussian-smoothed image. A comparison of these image reconstruction methods is show in \figref{reconstruct}. {\bf 3)} Applying our improved image reconstruction technique to the standard star data, we improved our photometric precision, resulting in a photometric agreement within 0.07~magnitudes between our \vp{} data and SDSS-g photometry (see \secref{photovalid}).

In this paper, we add two more wavelength bins to the data cube, one for the Swift UVOT UVM2 filter and one for the Spitzer IRAC 3.6\micron{} filter. Because SED modeling relies on the relative brightness at different wavelengths, it is essential that each spaxel in our final data cube samples the same part of the sky across wavelength and across instruments. Between these instruments, the poorest angular resolution comes from \vp{}, which is dominated by the effects of reconstructing the image plane from fiber spectra. As a result, it was necessary to degrade the UVM2 and 3.6\micron{} images to match the \vp{} data cubes to ensure a high fidelity comparison between the \vp{} spectra and the UVM2 and 3.6\micron{} photometry.

To do this, we sampled the UVM2 and 3.6\micron{} images with synthetic fibers, which were created using guide-star pointing data to match the actual \vp{} fibers during each exposure. We then reconstructed the UVM2 images and the 3.6\micron{} images using the same technique as with the \vp{} spectral cube. The results are shown in \figref{reconstruct}. These synthetic UVM2 and 3.6\micron{} images are exactly matched to the \vp{} spectral cubes in spaxel plate scale and sky alignment, and, since they were processed through the same image reconstruction algorithms as the \vp{} spectral cubes, they share any artifacts of the reconstruction process.

Finally, it is worth considering if variations in the PSF could affect our analysis. By chance, the PSF FWHM does not vary much between these instruments. More importantly, the blurring effects of the \arcsecond{4}{16} \vp{} fibers dominate over PSF, and, as described above, all images and spectra are subject to the same blurring due to coarse fiber sampling.

In \secref{analysis} we will discuss the methodology for fitting the star-formation history to each region in UGC~628. Since we have ensured that each spaxel in our data cube samples the same part of the sky, we are able to fit a unique history to each spaxel. Each pixel in the star-formation history maps presented below is individually derived from a spaxel in the data cubes.

\subsection{Photometric Validation}
\label{sec:photovalid}

\begin{table}
\caption{Whole-galaxy magnitudes compared before and after image reconstruction, along with known calibration uncertainties.}
\begin{tabular}{lccc}

Filter                        & Before  & After & Uncertainty\\
\hline
g (\vp{})                 & ---       & 15.792 & 0.07\footnotemark[1]       \\
g                         &  15.795   & 15.730 & 0.01\footnotemark[2]   \\
UMV2                      &  18.059   & 18.068 & 0.03\footnotemark[3]   \\
3.6\micron{}                &  15.389   & 15.352 & 0.016\footnotemark[4]   \\

\multicolumn{3}{l}{ $^1$ \footnotesize adopted as a result of this table}\\
\multicolumn{3}{l}{$^2$ \cite{Doi2010}}\\
\multicolumn{3}{l}{$^3$ Swift UVOT CALDB Release Note}\\
\multicolumn{3}{l}{$^4$ IRAC Data Handbook}\\

\label{tbl:photvalid}
\end{tabular}
\end{table}

As a test of our data reduction and image reconstruction methodology, we performed a series of comparisons between our data before and after reconstruction, and with SDSS DR14 broad-band data \citep{SDSSDR14}. To compare with SDSS data, we convolved the \vp{} spectral data cube with the SDSS \gband{} filter response function and reconstructed the field of view, as described above. The \gband{} filter is ideal for this comparison because it is entirely contained within the wavelength range of our \vp{} spectra. The first row in \tblref{photvalid} lists the magnitude of UGC~628 in the \gband{} image reconstructed from the \vp{} spectra.

We sampled the SDSS \gband{} image of UGC~628 with synthetic fibers and reconstructed the field of view, just as we did with the UVM2 and 3.6\micron{} images. The second row in \tblref{photvalid} lists the magnitude of UGC~628 in the SDSS \gband{} image before reconstruction (left) and after reconstruction (right). Likewise, the third and fourth rows in \tblref{photvalid} compare the magnitude of UGC~628 in the archival and reconstructed UVM2 and 3.6\micron{} images as a check on the reconstruction method and on the method used to sample the broad-band images.

The agreement between the before and after magnitudes is a test of the fidelity with which we can reconstruct the FOV from IFU data and the degree of uncertainty we introduce by doing so. Likewise, the agreement between the  the SDSS \gband{} magnitude and the \gband{} magnitude derived from our \vp{} spectra is a test of our data reduction and calibration techniques.

We see excellent agreement between the magnitudes before and after reconstruction. We conclude that sampling and reconstructing broad-band images does not impact our results.

The agreement between the SDSS image and our reconstructed \gband{} image is approximately 0.07~mag. This disparity is larger than our formal error bars, and larger than the 0.01~mag photometric calibration errors of the SDSS image \citep{Doi2010}. While SDSS images are typically too shallow for surface photometry on LSB galaxies, the galaxy-wide photometry has a much higher S/N, and should be considered reliable. Likewise, our \vp{} exposure times were calculated for S/N of ten at the \vp{} resolution of 5.1\AA, but averaging over the entire \gband{} boosts the S/N significantly, roughly $10\times$ excluding systematics.

If this disparity were an artifact of the image reconstruction process, then we would see a similar disparity between the before and after reconstruction magnitudes for the UVM2, 3.6\micron{}, and SDSS \gband{} images. Because these images were sampled with synthetic fibers, and those sampled values were processed in exactly the same manner as the \vp{} measurements, any photometric offsets introduced by the reconstruction process would be included in the difference between the magnitudes before and after reconstruction. Instead, these differences are all smaller than the 0.07~mag offset between the SDSS \gband{} and the \vp{} \gband{} photometry. Additionally, we tested the 2\arcsec{} kernel width used in image reconstruction by varying the kernel width, and found that a kernel width of 2\arcsec{} minimised the disparity between the before and after UVM2 and 3.6\micron{} magnitudes and between the \vp{} and SDSS \gband{} magnitudes.  We conclude that this disparity is likely a calibration error, either in our data or SDSS, and, to be conservative, we adopt the 0.07~mag disparity as our calibration uncertainty.

\section{Analysis}
\label{sec:analysis}

The goal of our analysis is to determine the likelihood distribution of past and present star-formation rates for each location within our target galaxy. To accomplish this, we divided the history of each spaxel, from 10~Gyr ago to present, into 15 semi-logarithmically-spaced age ranges, and used our data to determine the average star-formation rate within each timestep. The histories of all the spaxels collectively give a map of the star-formation rate in UGC~628 over cosmic time.

The number of timesteps is arbitrary, but it is limited by the sensitivity of the SEDs to age, our lack of knowledge of other parameters that influence the SED (see \secref{fixed}), and the quality of our data. Our model takes larger timesteps as we go further backward in time because the SED of a stellar population changes more slowly with age, making it more difficult to discriminate between populations of different ages. In this work the terms ``Gya'' and ``Mya'' are used to mean billions and millions of years ago {\it as seen from our perspective today}. Since UGC~628 is \about{78.3}Mpc away , it is necessary to add \about{243}~Myr of lookback time when comparing the timeline presented here to timelines for other galaxies.

For each of these age ranges, a normalised basis observation set (spectra and photometry) was generated from synthetic spectra produced with the \pegase{} spectral synthesis tool \citep{Fioc2011}. \pegase{} accepts a star-formation history as an input and produces a synthetic spectrum as an output, covering a wavelength range from the far UV to the far IR. The user is free to specify a parameterized star-formation rate (such as a single burst or an exponential decay), or to specify the exact star-formation rate at each point in time. The \pegase{} output lists luminosities for the nebular emission lines separately from the stellar continuum, leaving it up to the user to combine the two.

We start by creating basis observation sets by providing \pegase{} with a star-formation rate which is constant within each time step but zero at all other times. We then sample the output \pegase{} spectra to match the \vp{} spectra and convolve the \pegase{} spectra with the UVM2 and 3.6\micron{} filters to produce synthetic photometry, thereby creating synthetic spaxels identical in format and meaning to the spaxels in the real data cube (described in \secref{reduce}). These are the basis observation sets.

Each of these basis observation sets represents what our observations would be if given a spaxel within UGC~628 had only formed stars during a given window of time and at a rate of \sfr{1}. The process of SFH fitting then reduces to decomposing the real observations into a linear combination of these basis observation sets, convolved with an optimal set of environmental parameters, such as extinction normalization ($A_V$), choice of extinction law, and gas-phase metallicity. The extinction normalization is a free parameter in our fitting method, and is determined from the data on a spaxel-by-spaxel basis. The remaining environmental parameters are fixed; the values for these are discussed in detail in \secref{fixed}.

In addition to the rate of star formation in each epoch, the metallicity of the stars formed during each epoch also affects the observed spectra. One solution would be to adopt the current ISM metallicity that we reported in Paper~I, $\log{\rm O/H}+12=7.8$, or $\zstar = \zsol{13\%}$ \citep{Asplund2009}, however it is likely that the ISM metallicity has fluctuated over time due to enrichment and accretion of low metallicity IGM gas. Instead, we leave the stellar metallicity as a free parameter, and interpolate between four different metallicity tracks. The lowest track is around \zsol{3\%}, chosen to provide a robust lower bracket since it is much lower than expected for a galaxy of this mass, and lower than all but the most extreme dwarf galaxies \citep{Kniazev2018}. The second and third tracks, \zsol{30\%} and \zsol{60\%}, are more typical of what one would expect for an LSB spiral of this mass. In order to generously bracket the current ISM value and account for the possibility of more enriched stellar populations, we also include a fourth solar metallicity track. Our SFH fitting code generates stellar populations with intermediate mediate metallicities by interpolating between any two of these four tracks. Although the effects of metallicity on stellar spectra are, in detail, non-linear, the low-resolution of our \vp{} spectra makes our program fairly insensitive to these effects, allowing us to extract the star-formation histories. In Sections \ref{sec:uniqueness} and \ref{sec:stability} we present a thorough discussion of the robustness of our fitted histories. Nevertheless, we emphasise that the stellar metallicities presented should be taken as gross estimates only.  

It is worth noting that we have left the stellar metallicity as completely free parameter (within the bounds of \zsol{3\%-100\%}). It is not constrained to the gas-phase metallicity, and is allowed to go up or down with time.  This choice is reasonable given the growing body of evidence that galaxies may acquire pristine IGM gas after they have formed some fraction of their stellar population \citep[e.g.,][]{Huang2013,Sanchez2014,Elmegreen2016}. Such acquisition has the effect of diluting their ISM and necessarily lowering the metallicity of subsequently formed stars. For example, in their study of the stellar populations in the dwarf galaxy DDO~68, \cite{Annibali2019} find that the typical stellar metallicity is  around ten times as great as the gas-phase metallicity. DDO~68 shows signs of a recent merger with several smaller dwarfs, including a gas-rich dwarf, and the conclusion is that the intruder recently diluted the ISM. Although the stellar metallicity in most galaxies, including DDO~68, generally increases over time, any stars formed in DDO~68 in the near future will likely have a much lower metallicity that the older population in DDO~68.

The other factor which significantly motivates us to leave the stellar metallicity free to go up or down is that both the stars and the gas within UGC~628 have likely migrated over time. This is discussed further in \secref{results}, but the oldest populations are more smeared over the disk out than the younger populations. While this effect is primarily azimuthal, it is still possible for old populations with high metallicity to mix with young populations with low metallicity, even if the trend in metallicity over time for any one dynamical unit is purely monotonic. This consideration does not impact our ability to resolve the history of each spaxel, but our analysis cannot tell us exactly where the stars in each spaxel originated, an unavoidable uncertainty when studying the spatially resolved histories of galaxies.

\begin{figure}
  \includegraphics[width=\columnwidth]{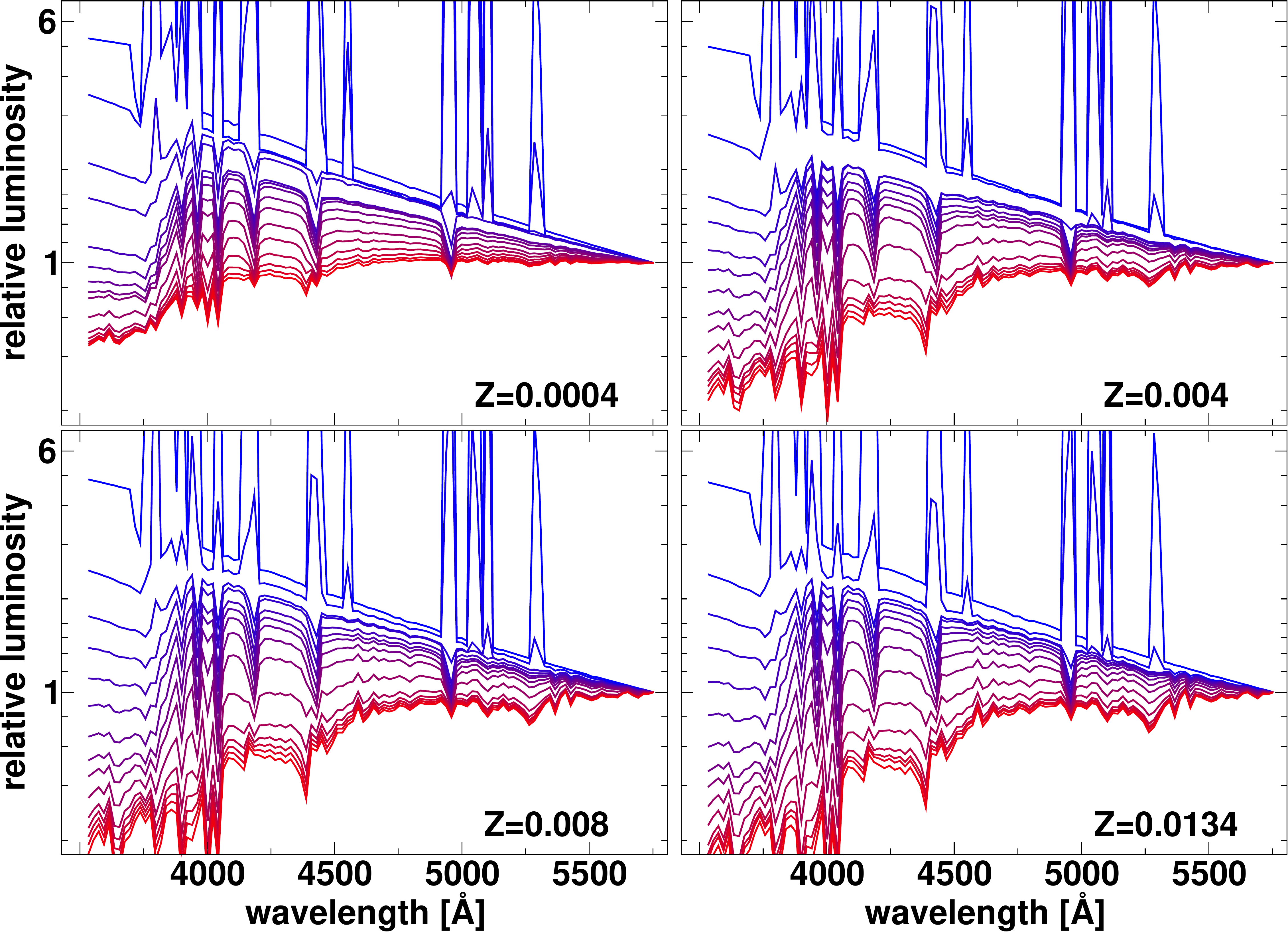}
  \caption{The synthetic \vp{} spectra in each timestep, grouped by metallicity. The units are relative luminosity density, normalised to the red end of the spectra. The youngest timesteps are blue, the oldest are red.}
  \label{fig:basis}
\end{figure}

For visual comparison, the synthetic \vp{} spectra from each timestep and each metallicity are shown in \figref{basis}. These are the spectral components of the basis observations. To guide the eye, the spectra are colour coded with the youngest timesteps coded as blue and the oldest coded as red. The spectra show the expected trends --- overall reddening with age, the disappearance of emission lines within the first few Myr, and, later, the strengthening of metal absorption lines with respect to the hydrogen absorption lines. Although less drastic that the effects of age, the differences between the spectra due to metallicity are significant, particularly at the blue end of the spectrum, underscoring the need to anchor SED fits with UV measurements.

Summing over bins in both time and metallicity, the standard $\chi^2$ comparison between model and observation can be expressed as:

\begin{equation}
  \chi^2 = \sum_\lambda\left[ \left( \frac{f_\lambda - \sum\limits_b w_b g_{\lambda,b}}{\Delta f_\lambda} \right)^2\right]
\end{equation}

where the $f_\lambda$ are the observations, $g_{\lambda,b}$ are the model basis spectra, and $w_b$ are the relative weights, determined from the star-formation rate and stellar metallicity at each timestep. By dividing each disparity term by $\Delta f$, this classic form has the desirable property that each term is inversely weighted by the uncertainty in the measurements and, assuming that $\Delta f_\lambda$ is of a similar order of magnitude as the disparity between the model and the data, each term represents a relative disparity between model and data that contributes equally to $\chi^2$ even if the $f_\lambda$ values vary strongly with $\lambda$.

Near the solution, this assumption is reasonable. However, any search algorithm must explore parts of parameters space far from the solution. In such cases an asymmetry manifests itself: in the case where the model values are too low, there is a maximum disparity that can be incurred (100\%), whereas for model values that are too large there is no maximum. As an example, consider a candidate solution for a region with bright emission lines in which the lines are fit correctly but the continuum is underestimated. Ideally the algorithm would raise the continuum and refit the lines, but because underfit values are penalized less than overfit values, this candidate solution is at a local $\chi^2$ minimum.

To create this symmetry and still retain the relative disparity aspect, instead of dividing each disparity term by $\Delta f_\lambda$ we divide by the geometric mean of the model and the data. To additionally capture the aspect of the classic form of $\chi^2$ wherein each term is inversely weighted by its relative uncertainty, we multiply each term by its signal-to-noise $f_\lambda/\Delta f_\lambda$. Near the solution, this reduces to the original $\chi^2$. This modification is purely to aid the search algorithm, and has no significant impact on the interpretation of the final $\chi^2$ values.

Additionally, in recognition of the fact that some of our wavelength bins are wide photometric bins ($\Delta\lambda > 1000\rm\AA$) while others are much smaller spectral bins ($\Delta\lambda = 20\rm\AA$), each disparity term is multiplied by the relative size of the bin $N_\lambda$.

\begin{equation}
  \chi^2 = \sum_\lambda \left[ N_\lambda  \frac{\left(f_\lambda - \sum\limits_b w_b g_{\lambda,b}\right)^2}{f_\lambda \sum\limits_b w_b g_{\lambda,b}} \left(\frac{f_\lambda}{\Delta f_\lambda}\right)^2 \right]
\end{equation}

Finally, for the sake of clarity, we rewrite our model spectra $g_{\lambda,b}$ as the product of model luminosities $L_{\lambda,b}$ and environmental factors $\phi_\lambda$, which include fixed parameters such foreground extinction and ISM metallicity, as well as the fitted model parameter $A_V$:

\begin{equation}
  \chi^2 = \sum_\lambda \left[ N_\lambda  \frac{\left(f_\lambda - \sum\limits_b w_b L_{\lambda,b}\phi_\lambda\right)^2}{f_\lambda \sum\limits_b w_b L_{\lambda,b}\phi_\lambda} \left(\frac{f_\lambda}{\Delta f_\lambda}\right)^2 \right]
\end{equation}

\subsection{Fixed Model Parameters}
\label{sec:fixed}

\begin{figure*}
  \includegraphics[width=0.9\textwidth]{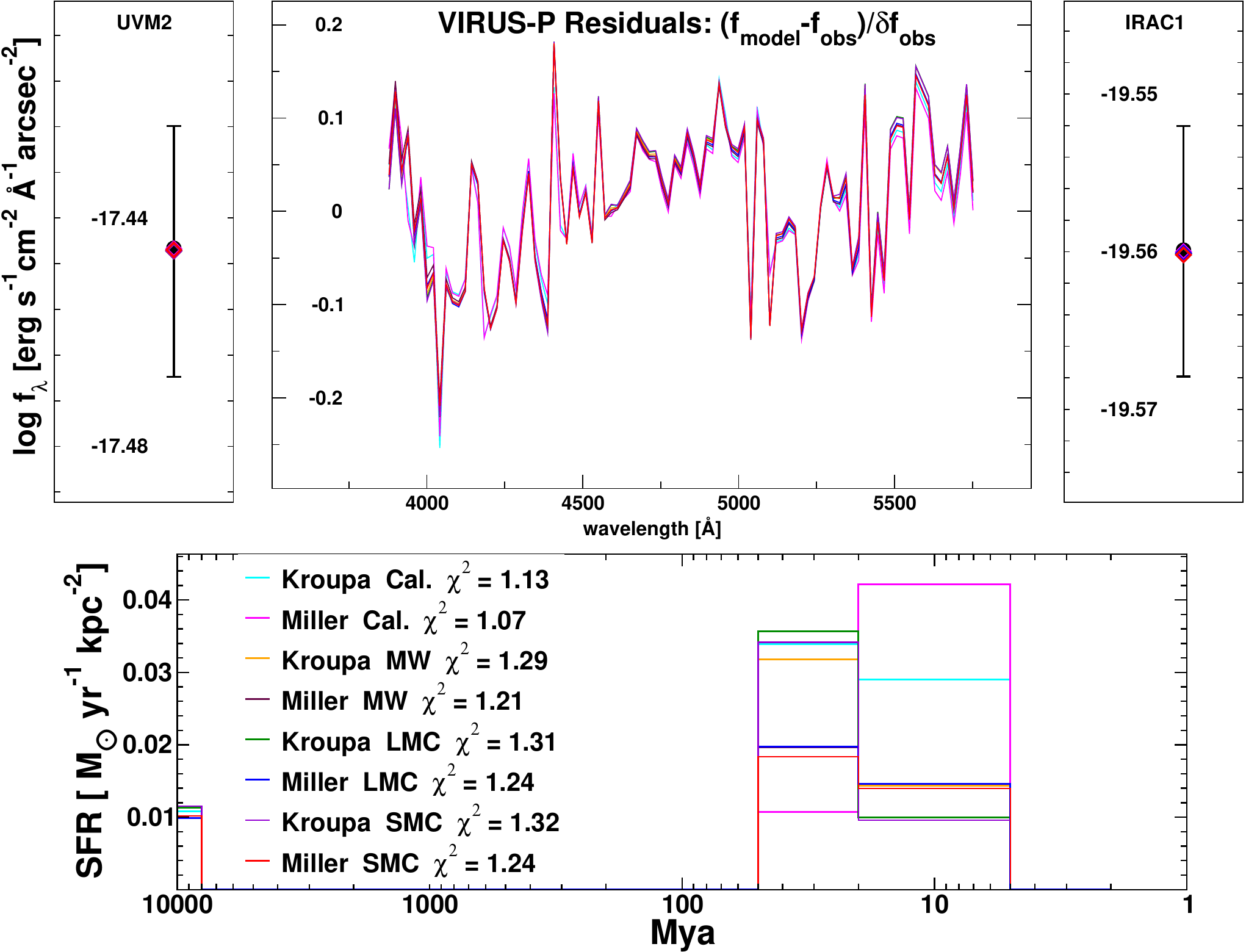}
  \caption{The best fitting photometry and residual spectra to the large star-forming complex identified as Region~$A$ using a battery of combinations of IMFs and extinction laws. Since the residual spectra are represented as fractions of the observed uncertainties, the $y$-axis of the top central panel ranges from $-20\%$ to $+20\%$ error. Region~$A$ was selected for this experiment because the bright emission lines, particularly \HB{}, provide a strong secondary diagnostic of the current and recent star-formation rate, and the effects of the choice of IMF and extinction law are more tied the current star-formation rate than the older population. The Calzetti Law is the best-fit extinction law, but the gross characteristics of the fitted history are not strongly affected the choice of IMF. The best overall fit is the Calzetti Extinction Law with the Miller-Scalo IMF.}
  \label{fig:perm}
\end{figure*}

The \pegase{} spectral synthesis tool provides the luminosities of the stellar continuum and emission lines separately. We chose to adopt the emission line luminosities produced by \pegase{}, with the exception of the \ionl{O}{2}{3727}, \ionl{O}{3}{4959}, and \ionl{O}{3}{5007} lines. Like many of the emission lines, these are sensitive to the ISM metallicity and ionization parameter, however unlike many of the other metal lines, they contribute a significant fraction of the the total luminosity in star-forming regions, especially the \ionl{O}{3}{5007} line. The ratios of these lines to \HB{} are related to the ISM metallicity and ionization parameters. In Paper~I we used observations of these lines in select \HII{} regions in UGC~628 to derive a metallicity and ionization parameter profile with the R23 bright line method. Our findings indicate that UGC~628 has a fairly flat metallicity and ionization profile, with values around $\log{\rm O/H}+12\approx7.8$ and $\rm U=-2.8$. Since there is comparatively little variation in these values (in the \HII{} regions where they can be sampled), in this paper we adopt these metallicity and ionization parameter values globally, even in locations where they cannot be measured with high fidelity. Our modeling program then corrects the fluxes of these oxygen lines so that their ratios against \HB{} are consistent with the adopted metallicity and ionization parameters, essentially the reverse of the process used for the emission-line bright regions in Paper~I.

However, we have fewer direct constraints on the stellar IMF and extinction law. The stellar IMF and choice of extinction law primarily affect measurements of the current/recent star-formation through their influence on UV continuum. Since emission lines, particularly \HB{}, also track the current star-formation rate, we have chosen to use the large star-forming complex on the North-Western limb of UGC~628, identified in Paper~I as Region~$A$, as a laboratory to study our choice of IMF and extinction law. As a case-study, we fit the brightest part of Region~$A$ with a combination of IMFs and extinction laws, permuting the IMFs from \cite{Kroupa1993} (top-heavy) and \cite{Miller1979} (bottom-heavy) with the extinction laws from \citep{Cardelli1989} (Milky Way Law), \cite{Koornneef1981} (LMC Law), \cite{Rayo1982} (SMC Law), and \cite{Calzetti1994} (starburst galaxies). This results in eight combinations. As noted in \secref{analysis}, the extinction normalization ($A_V$) is a free parameter in the fitting processes.

\figref{perm} presents the best fit histories for each of these eight models for Region~$A$, along with observed and model photometry and the residuals between the observed and model spectra. The normalised $\chi^2$ values for each model are also listed in the figure legend. All the models agree well with the observations, as can be seen from the fact that all the residual spectra in \figref{perm} fall entirely within the 20\% error bars and mostly within the 10\% error bars, similar to the 0.07~mag uncertainty we quoted in \secref{photovalid}. Because the residual spectra are also very similar to each other, and we conclude our methodology is not capable of determining which extinction law or which IMF is the appropriate choice. However, our goal is to determine the star-formation history, not the IMF or extinction law.

The fitted histories shown in \figref{perm} are all fairly similar to each other. All the histories have in common that Region~$A$ has an 8-10~Gyr old stellar population, and then a younger population that was formed between 5-50~Mya, with nearly zero current star formation and nearly zero star formation between 8~Gya and 50~Mya. Since the histories are similar, we conclude that the choice of extinction law and IMF does not significantly impact the gross characteristics of our results. However, since the differences between the histories in the 5-50~Mya range are roughly $\sim 2\times$, it is worth considering which extinction law and IMF to choose.

The two models with the lowest $\chi^2$ values are those that use the Calzetti Extinction Law; the Calzetti Law is naturally most appropriate for our project since  it is derived from unresolved stellar populations mixed with star forming regions. The choice of IMF seems to have less of an effect, however the bottom-heavy Miller-Scalo IMF is a slightly better fit. Hereafter, we adopt the best fit model, \cite{Miller1979} IMF and the \cite{Calzetti1994} extinction law, and note that this choice does not significantly impact our conclusions.\\

\section{Confidence Limits}
\label{sec:error}

The reconstructed star-formation history of UGC~628 is illustrated in \figref{timeline}. The top left panel shows the star-formation rates only, the top right panel shows the star-formation rates colour-coded by the stellar metallicities, the bottom left panel shows the confidence limits the star-formation rates, and the bottom right panel shows the total mass formed in each epoch. Because UGC~628 is an inclined disc, the overall shape in \figref{timeline} is an ellipse. The edges of the ellipse look tattered in many places; as the light profile tapers off, the S/N falls below the threshold where fitting is possible, leaving an uneven boundary. Several spots are masked out from our analysis due to contamination from foreground stars and a background galaxy. The implications of the star-formation history will be discussed in \secref{results}; here we discuss the calculation of the confidence limits and diagnostics of systematics.

\subsection{Limit Determination}
\label{sec:limit}

\begin{figure*}
 \includegraphics[width=.9\textwidth]{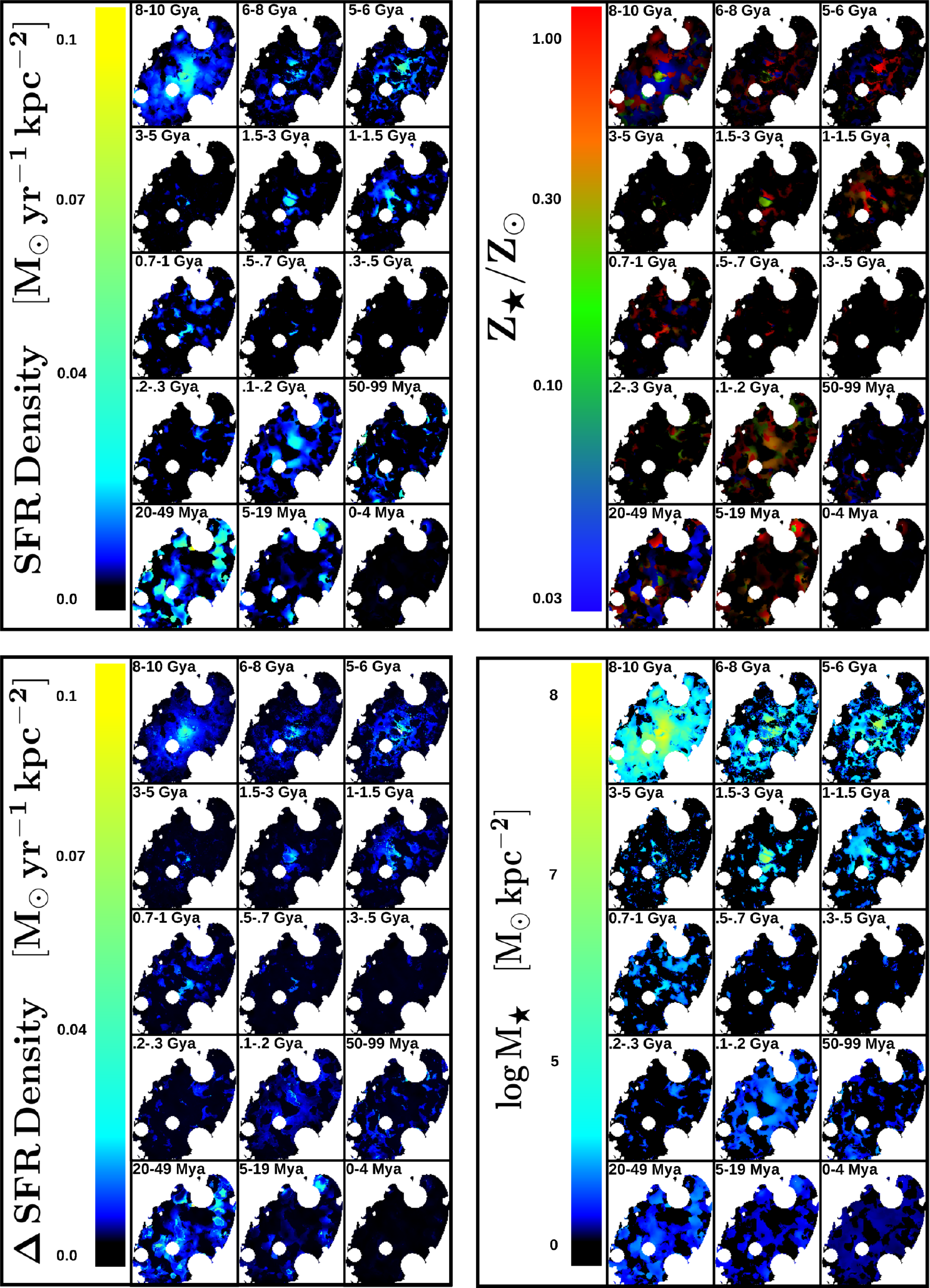}
  \caption{{\bf top left: } A map of star formation during 15 epochs in the history of our first target LSB galaxy. {\bf top right: } Like above, except points are colour-coded based on the metallicity of the stars formed during each era. {\bf bottom left: }Uncertainties in the star-formation rates shown in the top panel. The units and scale are the same as the top panel. {\bf bottom right: }The total stellar mass contributed to UGC~628 for formation in each epoch.}
  \label{fig:timeline}
\end{figure*}

In order to estimate the confidence limits on the star-formation rate in each epoch, we determined, for each pixel, the limits on the star-formation rate in each timestep (and $A_V$) that correspond to $\Delta\chi^2=\chi^2-\chi^2_0=1$. Many of the star-formation rates have a degree of covariance; that is, if the star-formation rate in one bin is changed, it may be possible to mitigate the effects on $\chi^2$ by shuffling that star-formation rate to an adjacent timestep. \secref{stability} and \secref{uniqueness} explore the impacts of this solution degeneracy on our conclusions. While we do not find that solution degeneracy affects our overall picture of the star-formation history of UGC~628, it is the primary source of uncertainty in our star-formation rates. In order to fully capture this effect while exploring the $\Delta\chi^2=1$ limits for each parameter, the remaining parameters are re-optimised at each step. The $\Delta\chi^2=1$ limit for each parameter is adopted as the uncertainty for that parameter.

In \figref{timeline} we see that the uncertainties range from around 0.005~$\rm M_\odot yr^{-1}kpc^{-2}$ near the edge of the galaxy to 0.05~$\rm M_\odot yr^{-1}kpc^{-2}$ near the centre.  The amounts to roughly 25\% uncertainties in timesteps with modest levels of star-formation. The notable exception is the low uncertainties in the 0-4~Mya timestep; current star formation is tightly constrained. The uncertainties are highest near the centre in the oldest timestep because most of the star formation near the centre of UGC~628 took place in the earliest timesteps, and, since older stellar populations have less of an effect on the present-day spectrum, the constraints are weaker. In contrast, Region~$A$, which has a very high star-formation rate in a recent timestep, shows very small uncertainties, even in the early timesteps.

\subsection{Linearity Validation}
\label{sec:uniqueness}
\begin{figure*}
  \includegraphics[width=0.9\textwidth]{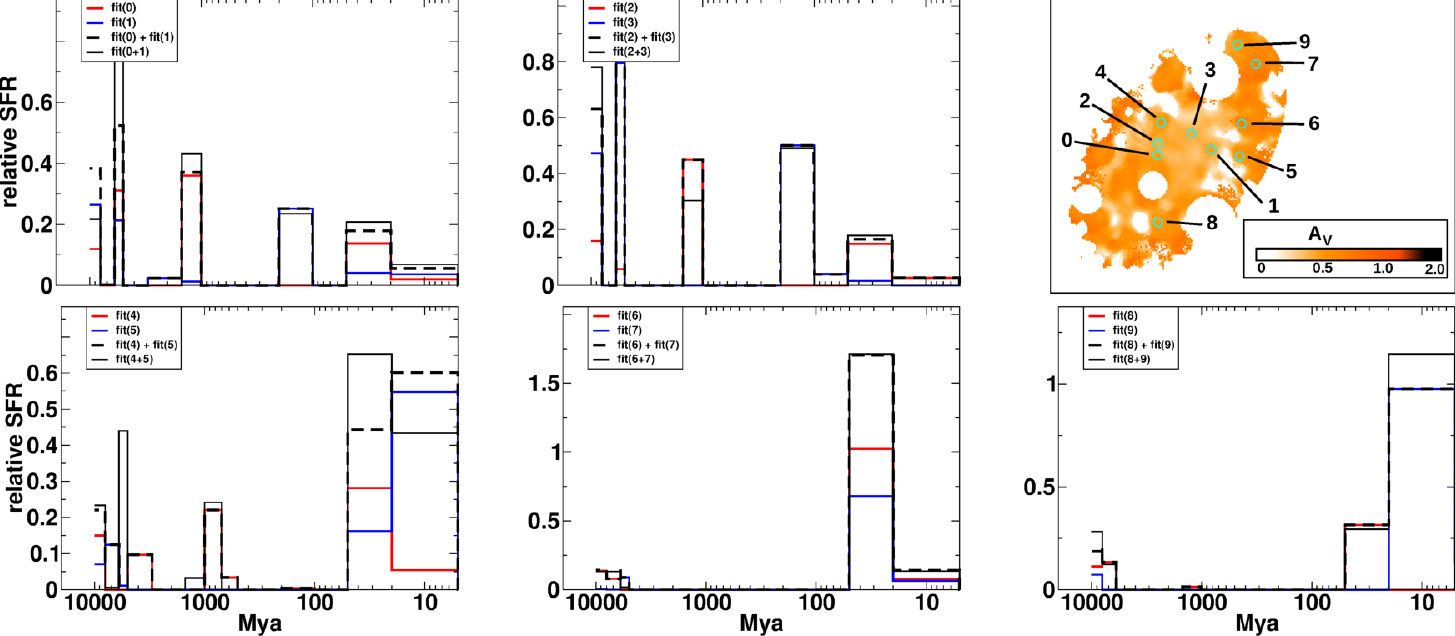}
  \caption{{\bf left:} Five pairs of regions in UGC~628 (top right), each selected to have identical $A_V$ values but different SF histories. For each pair, we show the fitted SFH for each separate region (coded red and blue), along with the sum of the two separate fits (black dashed line), and the fit of the summed SED of the the two regions (black solid). The similarity between the black dashed and black solid lines indicates our ability to determine the history of any composition stellar populations as a composite of their histories. The discrepancy between the back dashed and solid lines represents a systematic uncertainty in our methodology.}
  \label{fig:linearity}
\end{figure*}

Each resolution element in our SEDs will contain the light from different stellar populations with different histories blurred together. Ideally, we should derive the correct composite history from a composite population. A flawed methodology might produce a fitted SFH that disproportionately favors one sub-population, or, worse, produces a history that does not accurately represent any sub-population.

After the initial fitting was complete, we identified pairs of regions with similar best-fit extinction parameters but different histories, shown in \figref{linearity}. We then added their spectra and photometry to create summed SEDs, and fitted those summed SEDs with the same methodology that we applied to individual regions. If our methodology is robust, then the SFH fitted to the summed SED should be the sum of the fits to the individual SEDs. Note that it is important that we chose regions with similar extinction parameters since, unlike starlight, extinction does not add linearly.

We see in \figref{linearity} that the fitted SFHs of the summed SEDs closely resemble the sums of the fits to the individual SEDs. The match is not exact, but the gross characteristics of the SFHs match in all cases. We take this as a validation of our method, and a confirmation that we are able to spectroscopically distinguish multiple histories even when they are combined into a single SED. The small disparity between the sums of the fits to the individual regions and the fits to the summed SEDs can be considered a rough estimate on the confidence of our method, independent (though generally consistent with) the uncertainties presented in \secref{limit}.

\subsection{Numerical Stability Validation}
\label{sec:stability}
With the fitted star-formation histories in hand, we consider one final issue with our analysis: the possibility that the residuals between the model and the data have only a shallow dependence on one or more of the parameters, or perhaps a linear combination of parameters. For example, since the colours of older stellar populations change more slowly than the colours of younger populations, our residuals might depend on the total sum of star formation prior to 1~Gya, but with only a weak preference for the particular age range. In this scenario, a fitting algorithm might consistently converge on a solution for each region, when in fact a radically different solution is nearly as good a fit.

The natural method to test for cases where solutions have a shallow dependence on data is to refit the data with a small variance added; if the solution is robust, the solution will vary slightly, but if the solution has a shallow dependence on the data, the solution will vary wildly. The pixel-to-pixel variation of our best-fit star-formation histories, shown in \figref{timeline}, naturally provides this test.

In \secref{obs} we describe how our data cubes and SFH maps are rendered at a somewhat arbitrary plate scale of \arcsecond{0}{5}, significantly oversampling the PSF (dominated by the 2\arcsec{} seeing at Mt. Locke). Our adopted \arcsecond{0}{5} plate scale gives us the a final check on the stability of our method.

At the 78.3 Mpc distance of UGC~628, \arcsecond{0}{5} corresponds to 190pc. Star-forming regions and groups of stars may be smaller than this scale so, without atmospheric seeing, we might expect neighboring pixels in our SFH maps to have radically different histories. With the 2\arcsec{}, the light from each stellar population is spread across approximately four spaxels. If our algorithm works correctly, fitting each spaxel separately (without knowledge of its neighbors) should produce histories that vary smoothly from pixel-to-pixel, with no rapid variations on scales smaller than four pixels. If our algorithm is numerically unstable (as with parameter degeneracy), we might see significant pixel-to-pixel `speckling' in the SFH maps, caused by minor differences in the observed flux leading to significant changes in the derived history.

Examining the solutions presented in \figref{timeline}, the majority of the frames show very few cases of significant pixel-to-pixel variations in the SFHs.  Careful examination of the oldest bins reveals some speckling; these are the timesteps that are the closest to degenerate, since the composite spectra of older stellar populations change more slowly. Aside from this, the solutions vary continuously, gradually transitioning from one to the other, just as the values in the data cubes do. We conclude that our model and methodology are highly unlikely to suffer from serious issues with solution degeneracy, and are stable against small perturbations in the measurement values.


\begin{figure}
 \includegraphics[width=0.5\textwidth]{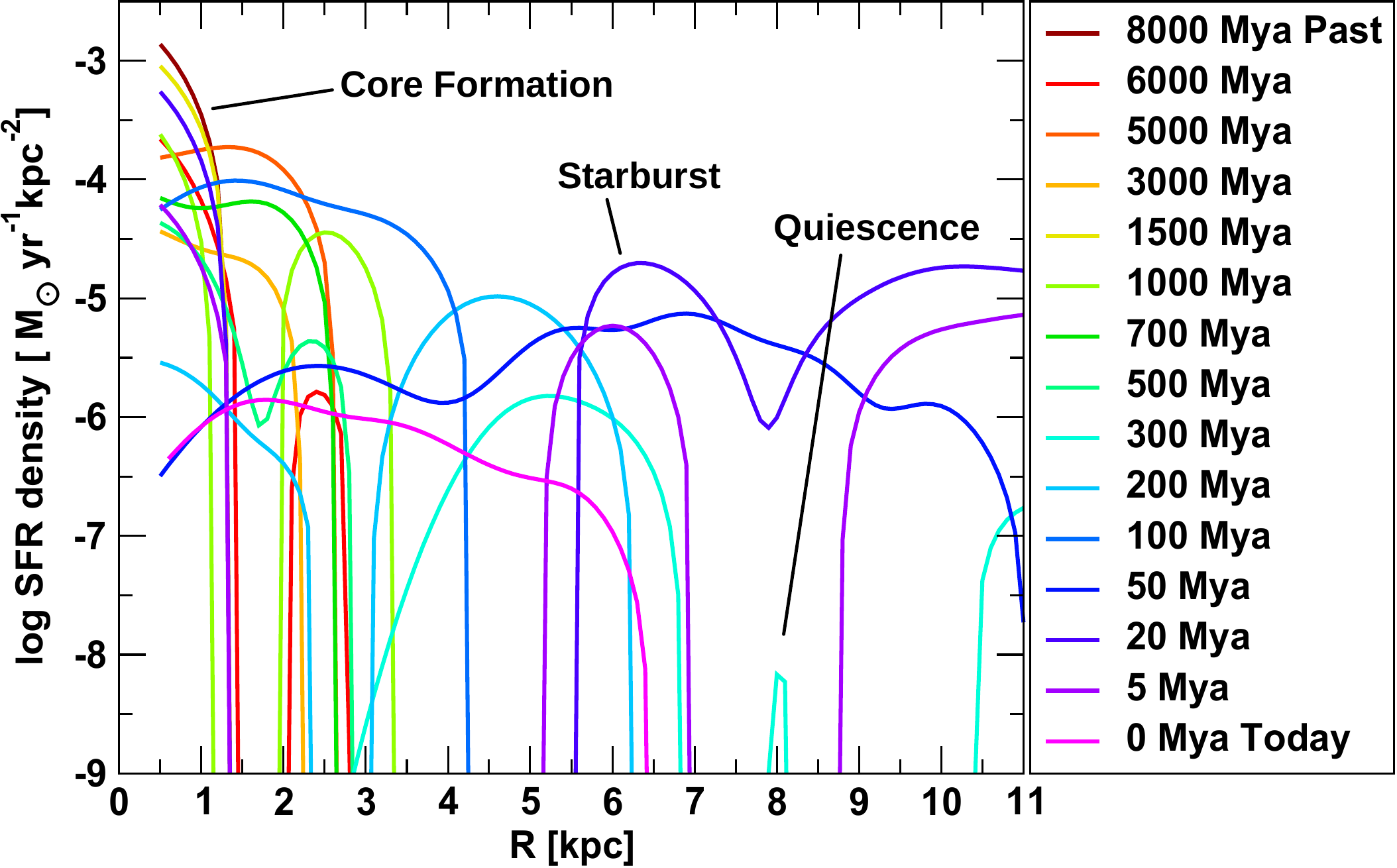}
  \caption{Azimuthally averaged star-formation rate densities, derived from the star-formation rate maps in \figref{timeline}. For the sake of clarity, the star-formation rate densities shown here have been smoothed with a Gaussian kernel with a FWHM of 0.5~kpc. These curves clearly illustrate a migration of centrally concentrated star formation early in the history of UGC~628 to disc-wide patchy star formation continuing through present-day.}
  \label{fig:radial}
\end{figure}

\begin{table}
\caption{Brief History of UGC~628}
\begin{tabular}{lllllr}
 era &  SFR & \ensuremath{\rm M_\bigstar} & \ensuremath{\rm\mu_0(B)} & \ensuremath{\overline{R_{SF}}}\\
 Mya & \ensuremath{\rm M_\odot\,yr^{-1}} & \ensuremath{\rm 10^{9} M_\odot} & & kpc \\
\hline\\

8000-10000 & 15.18 & 22.2 & 21.2 & 5.0 \\
6000-7999 & 3.29 & 23.7 & 22.4 & 5.9 \\
5000-5999 & 6.95 & 27.9 & 22.1 & 4.1 \\
3000-4999 & 1.01 & 27.1 & 22.6 & 4.8 \\
1500-2999 & 1.51 & 27.6 & 22.2 & 2.9 \\
1000-1499 & 4.65 & 29.2 & 21.9 & 4.1 \\
700-999 & 2.07 & 29.3 & 22.2 & 5.3 \\
500-699 & 0.54 & 29.2 & 22.5 & 7.7 \\
300-499 & 0.38 & 29.1 & 22.6 & 7.3 \\
200-299 & 1.52 & 29.1 & 22.7 & 7.2 \\
100-199 & 8.30 & 29.8 & 22.7 & 5.0 \\
50-99 & 4.67 & 29.9 & 22.7 & 8.3 \\
20-49 & 20.93 & 30.5 & 22.5 & 7.7 \\
5-19 & 5.52 & 30.5 & 22.7 & 9.1 \\
0-4 & 0.42 & 30.5 & 22.7 & 7.3 \\
\label{tbl:timeline}
\end{tabular}
\end{table}

\begin{figure}
 \includegraphics[width=0.5\textwidth]{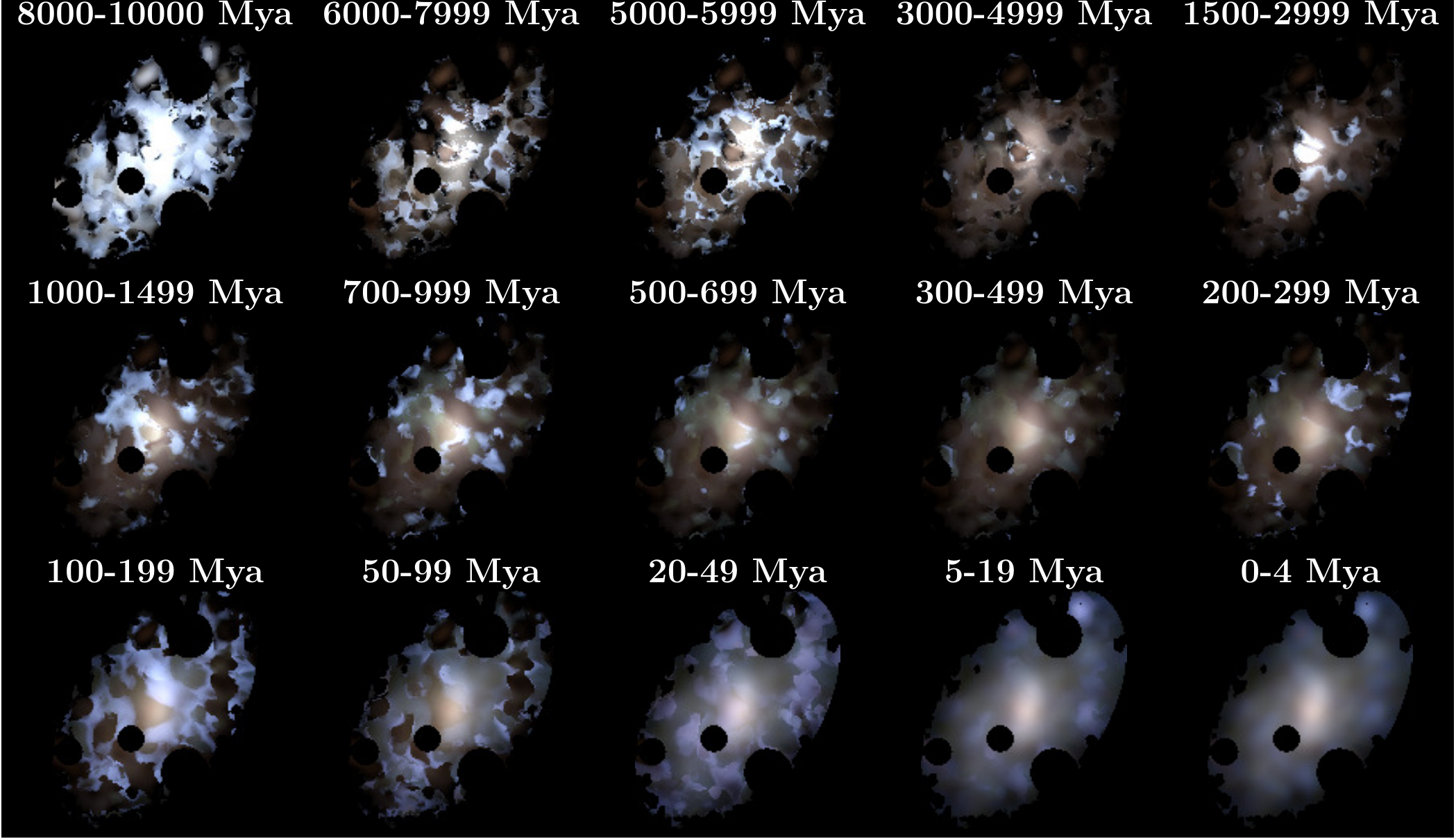}
  \caption{A UBV recreation of the stellar populations of UGC~628 as they were at the end of each epoch, assuming present-day extinction values. We have not corrected for orbital motion, and the stellar populations are represented at their current locations. The final frame is a match to the present-day appearance.}
  \label{fig:ubv}
\end{figure}

\section{Discussion}
\label{sec:results}

As mentioned above, the reconstructed star-formation history of UGC~628 is illustrated in \figref{timeline}. Azimuthally averaged histories are shown in \figref{radial}, and interesting galaxy-wide properties in each summarised in \tblref{timeline}. In order to compare to present-day galaxies, \figref{ubv} shows a recreated UBV image of UGC~628 for each timestep, assuming present-day extinction. Note that because of this assumption, the data in \figref{ubv} can only be used for gross characterization, and even then only with caution. Also note that \figref{ubv} is not a recreation of the actual appearance of UGC~628 during these epochs since we are not correcting for the orbits of the stars, and many dynamical times have passed since the earliest timesteps.

What follows is a discussion of the fitted history of UGC~628 set in the context of what is known about LSB spirals in general, both from other observations and from models. We emphasise that UGC~628 is only one object, and broader conclusions will require a larger sample. Using \figref{timeline} as a guide, we break the history of UGC~628 into several eras, summarised below:\\

\subsection{Timeline of Events}
\label{sec:timeline}

{\bf 8-10 Gya} Intense star formation is concentrated in the central 5kpc, but with noticeable levels of star formation through the disc. 60\% of the stellar mass was formed during this era (see bottom right of \figref{timeline} and \tblref{timeline}). The stars formed in the central region during this era are currently spatially coincident with the bar, and it seems likely that this is the era of bar formation (see discussion in \secref{context}). The distribution of star formation in UGC~628 during this era is grossly in-line with inside-out formation, and the buildup of a now-mature stellar population near the centre explains the current red-to-blue colour gradient reported in Paper~I. We also see the formation of solar-metallicity stars in the outer disk. The buildup of metals is not surprising given the duration of this era and the level of star formation, however the contrast with the low metallicity star formation dominating the central region is interesting, and will be discussed further in \secref{context}. During this era the unextinguished \sbB{} was bright enough that, had UGC~628 been observed, it might have been classified as an HSB spiral.

{\bf 5-8 Gya} In these two timesteps star formation persists through the disc of UGC~628, albeit at a much lower rate. We see one knot of star formation offset from the centre, at about the position where the bar currently ends.  Interestingly, \figref{timeline} shows that the stars formed in UGC~628 during this era have a higher metallicity than the stars formed near the centre during the 8-10~Gya peak-bar era. Although, as discussed in \secref{analysis}, our stellar metallicities should be taken as gross estimates only, this difference between these eras makes sense chronologically: the early initial burst locally enriched the central gas.

{\bf 3-5 Gya} UGC~628 is fairly quiet during this period, with an average star-formation rate of around \sfr{1}. We cannot rule out short high amplitude bursts, such as the one seen in the later 20-49~Mya timestep, but we can rule out the possibility that they were common. This was likely a {\it nearly} burst-free era. A possible analog to UGC~628 during this phase is UGC~8839, with a centre that is fairly bright but somewhat red for an LSB spiral, but an extremely faint blue extended outer disc.

{\bf 700 Mya - 3 Gya} These three timesteps show a burst of star formation. In the 1.5-3~Gya timestep we see no activity except in the core. This is the last point in the history of UGC~628 when the core is more active than the outer disk. In the 1-1.5~Gya timestep the core goes quiet and the disk becomes active. The activity in the inner disk is not clearly associated with the bar. During this era the unextinguished value of \sbB{\approx 22} would make UGC~628 a marginal LSB galaxy.

{\bf 200-699 Mya} These three timesteps are very quiet. The average level of galaxy-wide star formation is similar to today, around \sfr{0.4}. Like the 3-5~Gya era, this was likely a nearly burst free era. The scarce star formation during the 200-299~Mya timestep shows a strong preference for the outer disc of the galaxy. 

{\bf 100-199 Mya} This timestep shows another burst of star formation, this time predominately in the outer disk. \figref{timeline} shows a ring of star formation around the core. We caution against over interpreting this as a true dynamically driven ring. More likely, this is an extension of the patchy mode of star formation down to smaller galactocentric radii. It is worth noting, though, that star formation now clearly avoids the core. Given the sharp swings in star-formation rate in subsequent timesteps (see below), it is possible that this timestep was mostly quiet with a brief episode of intense star formation; only an average value is possible in our analysis.

{\bf 50-99 Mya} A fairly quiet era. Low-level star formation is now strongly biased toward the outer rim of the galaxy. Because our analysis necessitates that the earlier timesteps be larger, more recent timesteps may give us insight into the short-term variations that may have occurred during earlier timesteps. This timestep may be representative of inter-burst periods during the earlier timesteps.

{\bf 20-49 Mya} A major episode of patchy, nearly disc-wide, low metallicity star formation. Star formation again avoids the core/bar region entirely, and is, instead, spread along the outer rim of the galaxy. This edge-dominated star formation would seem to be the likely cause of the mostly flat and slightly inverted metallicity gradient we reported in Paper~I. This event may have been driven by tidal torques from an interaction, but the seemingly undisturbed morphology of UGC~628 makes a major merger unlikely. During this era UGC~628 was still an LSB spiral, however, had UGC~628 been observed during this epoch, it might have been excluded from LSB samples because of its bright knots. The average star-formation rate was \sfr{\approx 20}, the highest in our reconstructed timeline; UGC~628 would have been considered a starburst galaxy since, at this rate, as it could have assembled its entire stellar population in 1.5~Gyr. However, since this era only lasts 29~Myr, it only accounts for 2\% of the stellar population.

{\bf 5-19 Mya} This post-burst era sees UGC~628 with one knot of younger stars, Region~$A$, located on the Northwestern edge of the disc.  Region~$A$ itself shows a significant drop in star-formation rate since the previous era, and also an increase in the metallicity of stars forming now compared to those formed in the previous era; as with the core enrichment seen in the first few timesteps, this makes sense chronologically if Region~$A$ self-enriched during the 20-49~Mya era.

{\bf 0-4 Mya} Present-day UGC~628 sees star formation is largely shutdown, with only low levels in a few remaining areas such as Region~$A$. We see that the current phase in the life of UGC~628 is simply an inter-burst period, with relics such as Region~$A$, the inverted metallicity profile, and the redder bar hinting toward a more active past.

\subsection{Comparison with Other Observed and Simulated LSB Spirals}
\label{sec:context}

One of the most striking aspects seen in our fitted history of UGC~628 is the large swings in star formation. The recent history of UGC~628 seems to be made of long periods of quiescence punctuated by eras of star formation. During the most recent episode of star formation UGC~628's star-formation rate was approximately $10\times$ what would be required to assemble the current stellar mass in a Hubble Time. Conversely, during the period between 300-700~Mya the star-formation rate was fairly low, similar to the current level. This picture is similar to that portrayed in other works. Recall that \cite{Boissier2008} use Galex FUV-NUV colours to show that LSB spirals are likely in a post-burst phase; having lost their most massive stars, LSB spirals have weak emission lines and red UV colours, but are still optically blue. We see the extreme bursts of star formation discussed in \cite{Boissier2008} mirrored in our fitted history of UGC~628.

The metallicity of UGC~628 that we report in Paper~I, $\log{\rm O/H}+12=7.8$, or roughly $\zsol{13\%}$, is in the range that \cite{Gerritsen1999} suggest could suppress the cooling and formation of molecular clouds and trigger the patchy/sporadic behavior that we see. The star-formation rate fluctuations predicted in \cite{Gerritsen1999} are powers of ten lower in amplitude, and the quiescent periods are shorter than our timesteps are capable of resolving, however their galaxy is almost a factor of ten less massive, and \cite{Boissier2008} find that the burst amplitudes and quiescent durations seem to correlate with galaxy mass. Suppressed star formation via chronically low metallicity gas would be in-line with the idea of galactic ``overfeeding'' that we suggested as a possible explanation in Paper~I, a scenario in which a galaxy accretes pristine IGM too quickly for normal enrichment processes and becomes trapped in a low metallicity / low star-formation rate state.

As discussed in \secref{intro}, \cite{Vorobyov2009} use numerical hydrodynamic modeling of an LSB galaxy, and they find that their simulated galaxy exhibits bursts of star formation on timescales less that 20~Myr, similar to what we find for UGC~628. Star formation in the simulated galaxy peaks 2.6~Gyr after assembly (9.8~Gya), declining to a current value of \sfr{0.08}. Star formation in UGC~628 also experienced a peak around a similar time, with about 60\% of its stars formed during the 8-10~Gya timestep (see \tblref{timeline}). This is grossly in keeping with the concept of galactic downsizing, wherein both models and large surveys suggest that most of the star formation in ``typical'' galaxies in the mass range of UGC~628 should be at $z\sim1$ \citep[e.g.][]{Leitner2012,Moster2013,Behroozi2013}. However, since the 8-10~Gya era, the behavior of UGC~628 has departed from the model galaxy in \cite{Vorobyov2009}, with star formation not declining steadily but instead exhibiting wild swings. Indeed, a particularly vigorous episode of star formation occurred fairly recently in the 20-49~Mya time-step.

The bar may figure significantly into the overall history of UGC~628. \cite{Chequers2016} provides a thorough dynamical discussion of the formation of the bar in UGC~628 specifically, and find that their model best matches the current appearance of UGC~628 several Gyr after peak bar strength. However, the evolution in appearance after peak bar strength is fairly slow, and so \cite{Chequers2016} note that the window of time during which the appearance would match is fairly wide. Indeed, \cite{Mihos1997} find that, since LSB spiral discs are so stable, bars may be difficult to form but, once formed, are long lived.

In the earliest era, 8-10~Gya, we see star formation cover the area of the bar, almost entirely and almost exclusively. As we noted above, many dynamical times have passed, and all we can say is that most of the stars currently spatially coincident with the bar formed during this time period, and that comparatively few stars formed during this period which are not now spatially coincident with the bar. \cite{Chequers2016} find that the bar in their simulated galaxy begins to form when their galaxy is 3~Gyr old and peaks in strength at 5~Gyr. If we assume that the events in the 8-10~Gya timestep represent peak bar strength, that would marginally allow the formation of UGC~628's disc within the accepted age of the universe. It is much more likely that star formation in the 8-10~Gya timestep is associated with the era of bar growth.

\begin{figure}
 \includegraphics[width=0.4\textwidth]{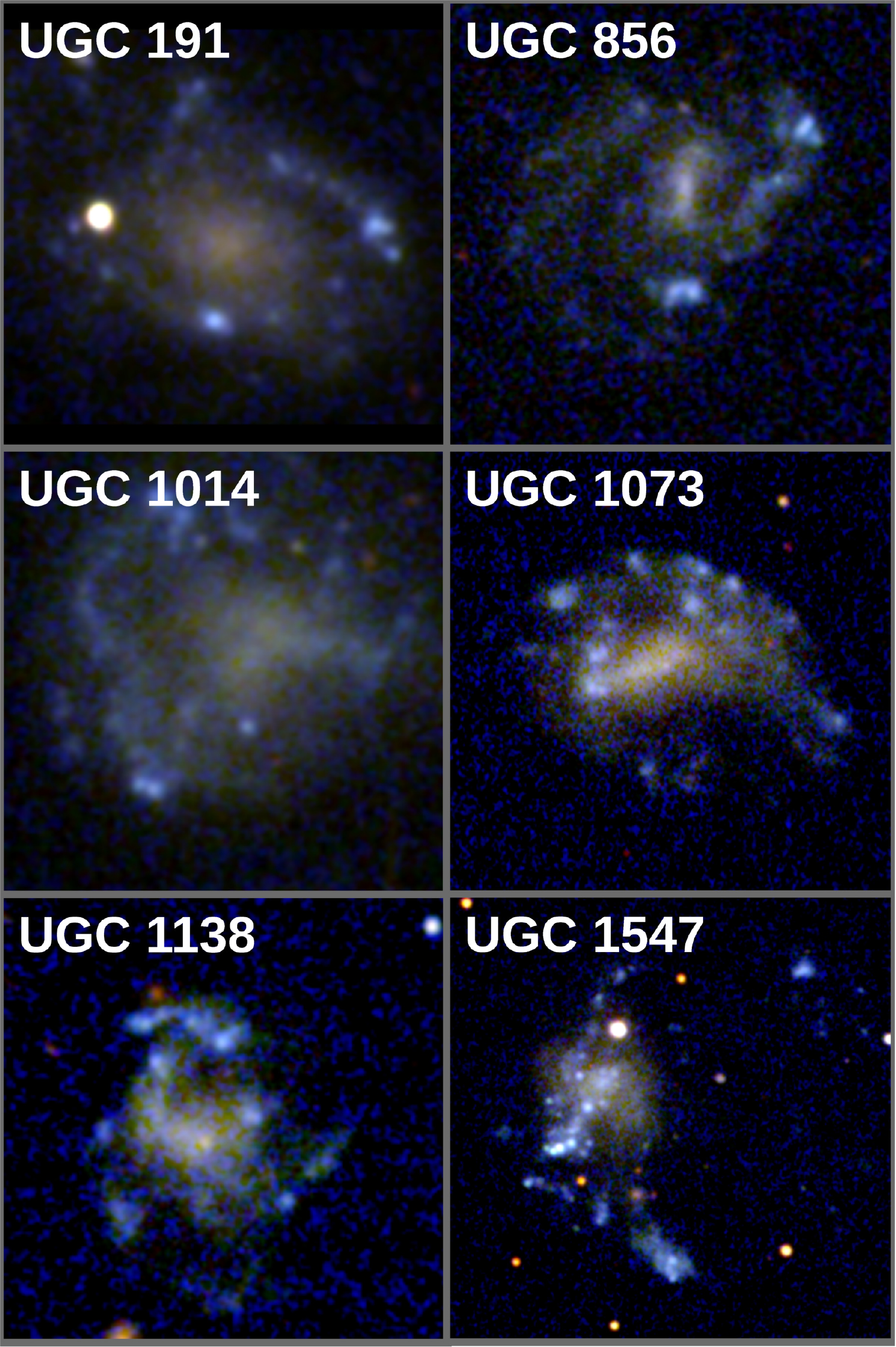}
  \caption{SDSS DR14 ugr images of size spiral discs galaxies with an overall low surface brightness but marked by large star-forming complexes in the outer disc. These may be an analogs to UGC~628 during the burst phases of its life cycle.}
  \label{fig:analogs}
\end{figure}

Regardless of the bar, early star formation in UGC~628 was much more centrally concentrated than it is now. In \tblref{timeline} and the lower right panel of \figref{timeline} we see that 60\% of the stellar population formed during this era, and within the central few kpc. This period of early formation matches the conclusion in previous works \citep[e.g.,][]{vandenHoek2000,Boissier2008,Schombert2014}, that, despite their overall blue colours, LSB spirals cannot be totally unevolved galaxies with only young stars. It would seem UGC~628 began its life on the inside-out track, and only after the shutdown of the bar/core did star formation follow the patchy/sporadic mode. The magnitude and concentration of early star formation sets the 8-10~Gya era apart from the others, and suggests that a real transition followed. Looking at the azimuthally averaged star-formation rate profiles in \figref{radial}, we see a clear migration of star formation out of the bar/core region. Likewise, in \tblref{timeline} the quantity $R_{SF}$ is the average radius weighted by the star-formation rate; from the earlier timesteps to the more recent ones it increases by almost a factor of two, and almost monotonically. Star formation in the most recent timesteps seems to preferentially avoid the bar/core, often occurring in clumps on the outer edge of the disc.

The evolution of the stellar metallicities shown in \figref{timeline} makes the galactic overfeeding scenario that we discussed in Paper~I seem unlikely, at least in UGC~628's centre. In the 8-10~Gya timestep we see the core dominated by low metallicity star formation while the outer disk hosted higher metallicity star formation. As with all our stellar metallicities, these should be taken with caution (see \secref{analysis}). But, even with significant systematic errors, it would still be true that most of the stars formed around the core had lower metallicities than those in the outer disk. This suggests that during this era the core had an ample supply of metal poor gas, possibly via accretion, while the outer disk was left to self enrich. In the next timestep we see an overall drop in star formation, with only higher metallicity stars forming around the core, as we would expect if the core had exhausted the supply of metal poor gas. If star formation in UGC~628 had been suppressed due to poor cooling or self shielding in low metallicity clouds, then we would expect to see star formation ``seeded'' in areas where higher metallicities are achieved, which is the opposite of our observations. As such, our data are more suggestive of shutdown due to gas depletion, expulsion, or heating.

During the 20-49~Mya era UGC~628 would have met both the technical definition starburst galaxy (star-formation rate high enough to grow its current stellar population in less than a Hubble Time), and of an LSB galaxy (faint central brightness). This statement is robust: even if our estimates for the star-formation rate are incorrect by a factor of nine, UGC~628 would still have been a starburst galaxy during this era. With a small number of extremely active star forming complexes on the outer edge of an otherwise faint disc, such a starburst LSB galaxy would differ significantly from the more traditional nuclear starbust galaxies.

During a starburst phase, especially 20-49~Mya, UGC~628 likely resembled the galaxies shown in \figref{analogs}. These galaxies would not normally be classified as LSB spirals because of their bright blue clumps, despite an otherwise low surface brightness disc. The core region in these galaxies is brighter and redder and the surrounding disk, though still comparatively faint and blue, similar to UGC~628. It is tempting to think of these objects as analogs to the Cartwheel Galaxy, which was likely a giant LSB disc before a face-on interaction with a smaller galaxy which plunged through the centre of the disk. However, like UGC~628, none of the examples in \figref{analogs} is a giant LSB, nor do any show signs of interaction or disturbance or have any obvious companions. In the case of UGC~628, the scenario where a dwarf galaxy plunges through the centre of the disk face-on is clearly ruled out as an explanation for the edge-dominated star formation since, in this scenario, we would expect star formation to start from the centre and work outwards, whereas we see random patches of star formation that simply tend to be found more often near the edge. We can also rule out a massive but distant companion since there are no galaxies near UGC~628 visible in the SDSS images.

If the recent starburst was due to an interaction, it was likely a small companion which has since merged with or been obscured by UGC~628. The upper right panel of \figref{timeline} shows that the stars formed during the starburst had a lower metallicity than most of the stars formed since the early HSB-like burst 8-10~Gya. One interpretation would have a small gas-rich galaxy merging with UGC~628, diluting the ISM metallicity, and triggering a starburst. However, this scenario would have to be reconciled with the flat/inverted gas-phase metallicity profile we reported in Paper~I. Simplistically, the addition of low metallicity gas at the edge of the disk would make the metallicity profile steeper. The enrichment from the starburst mitigates the problem, but it may not solve it entirely. This problem would be solved if the inflow of gas was distributed throughout the disk of UGC~628, but in that case it is unclear why the starburst was localized near the edge. 

It is also possible that the hypothetical companion was gas poor and/or did not contribute much to the ISM gas involved in the associated starburst. This scenario is particularly interesting because, if correct, it would lend credence to the concept that minor interactions and mergers with may play a significant role in the star-formation histories of their larger partners through means other than the simply contributing to the supply of cool gas. For example, \cite{Tanaka2017} find low surface brightness features around the Seyfert galaxy M77, which suggest a minor merger on a timescale compatible with the idea that the merger initiated a burst of star formation and potentially AGN activity.  Several works \citep[e.g.,][]{MartinezDelgado2012} find compelling evidence that activity in  the starburst dwarf irregular NGC4449 was triggered by a merger with an even smaller dwarf. At the extreme, even tiny galaxies such as DDO~68 show signs of starburst triggered by the accretion of faint dwarfs \citep{MartinezDelgado2012,Annibali2016}. In all of these cases, the tell-tail signs come from ultra-low surface brightness features, which would not be visible in any existing images of UGC~628. One interpretation of the star-formation histories presented in this paper is that we are seeing the accretion history of UGC~628, marked by edge-dominated star-forming events.

The bottom right panel of \tblref{timeline} and \figref{timeline} give a sense of the significance of the later bursts of star formation to UGC~628. 60\% of the stellar mass of UGC~628 was formed in an inside-out, HSB-like event. The later starbursts contributed 40\% of the stellar mass, but primarily much further out in the galaxy. Those characteristics are in line with what would be expected from accretion events. However, the accretion of stars normally contributes to the buildup of a spheroid, and these events seem to have contributed to the buildup of the disk. If they were accretion events, it seems likely that either they primarily contributed gas or their primarily influence was to disturb the existing gas disk.

We see evidence for annular morphology reminiscent of that found in UGC~628 in \HI{} observations of other LSB spirals. For example, \cite{Pickering1997} find large amounts of diffuse \HI{} in their sample of four giant LSB galaxies. The \HI{} is below the critical density for star formation \citep{Kennicutt1989} throughout most of the disc, reaching the threshold only in a few locations at large radii. If UGC~628 is analogous, \HI{} maps of UGC~628 would likely show a gaseous disc which is subcritical except where we see star formation, near the visible edge of the disc. In a more theory-oriented work looking at multi-component models of Milky Way-mass LSB spirals, \cite{Garg2017} suggest that disc stability in many LSB spirals may reach a local minimum and achieve star formation only at radii several kpc from the galactic centres. It seems plausible, then, that UGC~628 and the galaxies in \figref{analogs} represent different phases of internally driven disc evolution which, in these cases, favors star formation at large radii. 

Careful studies of HSB spirals show that edge-dominated star-formation and starbursts centered in the outer disk are, in fact, not unheard of in late-type spirals. \cite{Perez2013} examine the first 105 galaxies in the CALIFA survey, and find that inside-out formation is the dominant mode of galaxy assembly in systems larger than \mstar{=6\times 10^{10}\rm M_\odot}, but, below that limit, star formation is more evenly distributed throughout the disk, and small systems even exhibit outside-in formation. \cite{Huang2013} use spectra of 1000 galaxies to show that edge-dominant starbursts are more common in late-type and/or low surface density galaxies. These findings set a precedent for edge-dominated starbursts, although UGC~628 is more extreme than these examples. The outer-disk starbursts described in \cite{Huang2013} typically only account for 10-20\% of the total stellar mass, but 40\% of the stellar mass in UGC~628 formed in the patch/sporadic, edge-dominant mode.


We can also look to the gas-phase metallicity for clues to past star formation. In a study of star-forming galaxies with \lmstar{>10}, \cite{Moran2012} find that the most gas-rich 10\% of their sample show sudden drops in the gas-phase metallicity in their outer disks, which they interpret as a sign of recent accretion. In Paper~I we reported the opposite in UGC~628, a nearly flat but slightly inverted gradient in the gas-phase metallicity, as sampled by the R23 bright line method. When compared to the findings in \cite{Moran2012}, a simplistic interpretation would favor secular disk instabilities instead of accretion events as the drivers of the outer-disk starbursts in UGC~628. However, \cite{Bresolin2015} find that seemingly flat or shallow metallicity gradients in LSB spirals are actually very similar to the steeper metallicity gradients in HSB spirals when recalculated as gradients with respect to the exponential scale radii instead of physical units (kpc). This makes intuitive sense: if star formation in LSB spirals is spread out over a larger area, the enrichment will be as well. In this view, edge-dominated star formation would cleanly explain the larger scale radii of LSB spirals. If UGC~628 fits into this scheme in a straightforward way, it may be a typical object at a transitional time: The recent and fairly intense burst of star formation 20-49~Mya may have temporarily left it with a slightly inverted gradient, to be flattened out over time by further bursts and, to some extent, by radial migration. These hypotheses could be explored by metallicity measurements of the extended disc, beyond the bright \HII{} regions we report in Paper~I. 

\section{Summary and Conclusions}
\label{sec:summary}

By analyzing space-based UV and IR photometry in tandem with ground based IFU spectra we have produced spatially resolved maps of the star-formation history of the LSB spiral UGC~628. We have demonstrated that this technique is stable and reliable, and not particularly sensitive to most assumptions about environmental parameters. The exception is the choice of extinction law; after exploring a battery of extinction laws, we find that the Calzetti Extinction Law best agrees with our data. We find that, although UGC~628 exhibits only low-level star formation today, it has been more active in the past, broadly in agreement with existing numerical and hydrodynamic simulations. Our main findings regarding the history of UGC~628 are summarised below:\\

\begin{enumerate}
  \item UGC~628 is \about{10}Gyr old, with \about{60\%} of its stellar mass formed within the first 2~Gyr.
  \item Early star formation seems to have followed the inside-out pattern, as in HSB galaxies, and may be associated with the formation of UGC~628's bar.
  \item During the earliest era, 8-10~Gya, UGC~628 was likely an HSB spiral. Since that time, it has been an LSB spiral, although at times only marginally.
  \item After the first few Gyr, UGC~628 transitioned to a patchy/sporadic mode of star formation.
  \item Recent star formation has avoided the central 2kpc and, instead, shows a marked preference for the visible edge of the galaxy. It may be that these edge-dominated bursts mark the accretion of smaller objects.
  \item During some eras, such as 20-49~Mya, UGC~628 would have exhibited large star forming complexes on the visible edge of the galaxy. These properties would likely have excluded UGC~628 from most LSB catalogs and surveys, even though the underlying stellar disc would still have had \sbB{> 22}.
  \item During the 20-49~Mya era specifically the edge-dominated star-formation rate was high enough to qualify UGC~628 as a starburst galaxy.
  \item The current era can best be described as post-burst, with Region~$A$ as a relic from the recent episode.
\end{enumerate}

\subsection{Future Work}
\label{sec:futurework}

The recent burst in UGC~628 20-49Mya suggests the possibility of an interesting class of galaxy, a starburst LSB spiral. Further exploration into galaxies such as those shown in \figref{analogs} may reveal if they are truly analogs to UGC~628 during this period. If this patchy/sporadic mode of star formation is common among LSB spirals, then it may be possible to examine galaxies at different phases in the burst cycle to better explain the physics driving and inhibiting star formation in LSB spirals.

As before, we emphasise that UGC~628 is only one object, and we caution against drawing global conclusions of LSB spirals based on the history of UGC~628. Indeed, since it is a barred LSB spiral it may not even be typical for this class. Following up on our analysis of the star-formation history of UGC~628, the goal of the MUSCEL program is to extend our methodology to other LSB spirals. 

Preliminary analysis of LSB spirals in our sample suggests that comparatively flat ISM metallicity gradients are common in our sample, in keeping with earlier findings of other LSB galaxies \citep{deBlok1998I}. As noted in \secref{context}, the analysis in \cite{Bresolin2015} cleanly explains this in the context of enrichment: typical LSB spirals have flatter metallicity profiles because the stellar population have shallower profiles. UGC~628 would seem to be a transitional object, with a slightly inverted metallicity profile as a result of its recent burst of annular star formation. If this is correct, then we can utilise our program's ability to detect past bursts of star formation and see if there is a relationship between the ISM metallicity gradient and the location of and time since the most recent burst.\\

This work was done in collaboration with the Mount Holyoke College GeoProcessing Lab.\\

This work was performed in part using high performance computing equipment obtained under a grant from the Collaborative R\&D Fund managed by the Massachusetts Technology Collaborative.

\bibliographystyle{apj}
\bibliography{lsb}

\end{document}